\documentclass[aps,prb,preprint,superscriptaddress]{revtex4}

\usepackage{amsmath}

\begin{document}

\title{The Gibbs paradox and the distinguishability of identical particles}

\author{Marijn A. M. Versteegh}
\affiliation{Institute for History and Foundations of Science, Utrecht University, P.O.\ Box 80 010, 3508 TA Utrecht, The Netherlands} \affiliation{Debye
Institute for Nanomaterials Science, Utrecht University, Princetonplein 1, 3584 CC Utrecht, The Netherlands}

\author{Dennis Dieks}
\affiliation{Institute for History and Foundations of Science, Utrecht University, P.O.\ Box 80 010, 3508 TA Utrecht, The Netherlands}


\begin{abstract}
Identical classical particles are
distinguishable. This
distinguishability affects the number of ways $W$ a macrostate can
be realized on the micro-level, and from the relation $S=k \ln W$ leads to
a non-extensive expression for the entropy. This result is usually
considered incorrect because of its inconsistency with
thermodynamics. It is sometimes concluded from this inconsistency that identical particles are fundamentally indistinguishable after all; and even that quantum mechanics is indispensable for making sense of this. In contrast, we argue that the classical
statistics of distinguishable particles and the resulting
non-extensive entropy function are perfectly acceptable from both a
theoretical and an experimental perspective. The inconsistency
with thermodynamics can be removed by taking into account that the
entropy concept in statistical mechanics is not completely
identical to the thermodynamical one. We observe that
even identical quantum particles are in some cases
distinguishable, and conclude that quantum mechanics is irrelevant
to the Gibbs paradox.
\end{abstract}

\maketitle

\section{Introduction: The Gibbs Paradox}

Imagine two gas-filled chambers of the same volume, separated by a
partition. Both chambers contain equal amounts of the same gas in
equilibrium, consisting of the same number $N$ of identical
classical particles with the same intrinsic
properties. By ``identical'' we mean that the particles have equal intrinsic properties such as charge and mass. Both parts have the same total energy,
temperature $T$, and pressure. The partition is suddenly
removed. What happens to the entropy?

According to thermodynamics the entropy remains the same, because
the macroscopic properties of the gases in both chambers do not
change. From a thermodynamic point of view, which means restricting
ourselves to the consideration of macroscopic properties, nothing
happens. If $A$ is the macrostate with the partition in
place and $B$ the macrostate without it, the entropy difference is
defined as
\begin{equation}
S_{B}-S_{A}=\!\int_{A}^{B} \frac{dQ}{T},
\end{equation}
where $dQ$ is the heat transfer during a quasistatic process from
$A$ to $B$. Because the gases remain in equilibrium, the removal
of the partition is a quasistatic process. We have $dQ=0$, and hence $S_{A}=S_{B}$.

In statistical mechanics the entropy is usually taken to be
proportional to the logarithm of the number $W$ of microstates that are
compatible with a given macrostate: $S = k \ln W$, where $k$ is
Boltzmann's constant. When the partition is removed, the number of
available states $X$ per particle doubles: each particle now has
twice as much phase space available to it than it had
before.\cite{number of states} Hence, the multiplicity increases from $W_A=X^{2N}$ to $W_B=(2X)^{2N}$, which corresponds to an
entropy difference $S_{B}-S_{A} = 2kN \ln2$. This is the Gibbs
paradox: The entropy increases according to statistical mechanics
but remains the same in thermodynamics.

A traditional way of resolving this paradox is by denying that
permutation of identical particles leads to a different state. The
real multiplicity is accordingly a factor of $N!$
smaller for a system of $N$ identical particles than what we
supposed previously. By the removal of the partition the
multiplicity now goes from $W_A=X^{2N}/(N!)^2$ to
$W_B=(2X)^{2N}/(2N)!$. With the help of Stirling's approximation it
follows that, in the thermodynamic limit $N \rightarrow \infty$,
$W_B=W_A$, which removes the discrepancy between statistical
mechanics and thermodynamics.

However, classical particles are always distinguishable by
their positions and trajectories, and thus in classical statistical
mechanics there appears to be no reason for the division by $N!$. According to several authors, quantum theory is therefore needed for justifying this solution of the Gibbs
paradox (see, for example, Refs.~\onlinecite{schrodinger 1948, huang
1963, wannier 1966, sommerfeld 1977, schroeder 2000, ben-naim
2007}). Identical quantum particles seem
indistinguishable from the start, because
quantum states of systems of identical particles must either be
symmetrical under permutation (bosons) or anti-symmetrical
(fermions), so that the exchange of particles leaves the state
invariant apart from a global phase factor and the multiplicity
$N!$ never enters.

If this reasoning were correct, then the Gibbs paradox would imply
that the world is quantum mechanical. It is difficult to
believe that a thought experiment from classical
physics could produce such a profound insight. Unsurprisingly therefore, doubts have been expressed concerning the
traditional solution of the paradox. For example, some authors
have argued that identical classical particles are also
indistinguishable, which would justify the factor $1/N!$ without recourse to quantum mechanics.\cite{hestenes 1970, fujita 1991,
nagle 2004, saunders 2006}

In this paper we follow an argument that is closer to the spirit
of classical mechanics. We accept that
identical classical particles are distinguishable and
that permutation of two of them leads to a different microstate.
Nonetheless, we shall show that the Gibbs paradox can be resolved within
classical physics. The key to the resolution is the
recognition that the entropy concept in thermodynamics is not
identical to that in statistical mechanics. This
observation will turn out to be sufficient for the resolution of
the paradox: neither the indistinguishability of identical particles
nor an appeal to quantum theory is needed. On the contrary, even identical quantum particles can sometimes
be distinguishable in the sense that is relevant here, so that
indistinguishability cannot be relevant to the Gibbs paradox even
in quantum mechanics.

\section{Permutations of identical classical particles}

Classical particles, paradigmatically represented by impenetrable
spheres, are the example par excellence of distinguishable
particles. At each instant of time they can be
individually labeled by their different positions. Classical particles follow continuous and non-intersecting
trajectories in space-time. Therefore classical particles can be identified over time by their different histories.

Permutation of two identical classical particles produces a
different microstate. Imagine a situation in which there
is one particle at position $x_1$ and one at position
$x_2$. At a later instant there is again one particle
at $x_1$ and one at $x_2$. Assume that their respective
momenta are the same as before. What has happened in the meantime?
There are two possibilities: either the particle that was first at
$x_1$ is again at $x_1$ and the particle that was first at
$x_2$ is later again at $x_2$, or the particles have exchanged
their positions. The latter case would be different from
the former one, because it corresponds to a different physical process.
Although it is true that the two final two-particle states cannot be distinguished on the
basis of their instantaneous properties alone, their different
histories show that the particle at $x_1$ in one final situation
is not equal to the particle at $x_1$ in the other final
situation.

Given these observations, how can it be that some authors deny that identical classical particles can be
distinguished and maintain that permutation does not give rise to a different
microstate? One reason is that there is an ambiguity in the
meaning of the terms ``distinguishable'' and ``permutation.''
Consider the following statements: ``Two particles are
distinguishable if they can always be selectively separated by a
filter;''\cite{hestenes 1970} ``Two particles are distinguishable
if they are first identified as 1 and 2, put into a small box,
shaken up, and when removed one can identify which particle was
the original number 1.''\cite{nagle 2004} These
definitions of distinguishability indeed imply that identical classical particles
are always indistinguishable. The concept of
``permutation'' can similarly be understood in a way that is different from what we have described. Consider again
the microstate of two identical particles, one at $x_1$ and
another at $x_2$. If the particle at $x_2$ were at $x_1$ instead,
and the particle at $x_1$ were at $x_2$, with all properties
interchanged, there would be no physical differences, neither from
an observational point of view nor from the viewpoint of theory.
It is therefore reasonable to say that the two
situations are the same (see, for example, Ref.~\onlinecite{fujita 1991}).

This is a different kind of permutation than the physical
exchange we considered previously. In our first example the particles
moved from $x_1$ to $x_2$ and \textit{vice versa}.
Trajectories in space-time connected the initial state to the
permuted state. In contrast, in the alternative meaning of
permutation, the exchange is not a physical
process. Instead, it is an instantaneous swapping that
occurs in our imagination, and thus it exchanges nothing but indices and does
not need trajectories.

A third sense of permutation is used
by Saunders.\cite{saunders 2006} One particle follows trajectory
1 and the other follows trajectory 2. Imagine that the particle that
actually follows trajectory 1 instead followed trajectory 2 and
\textit{vice versa}. That would result in exactly the same
physical situation. As before, this notion of permutation involves the consideration of
states before and after the permutation that are not connected by
a physical process. A permutation in this sense occurs
in our imagination and exchanges an abstract identity (represented by particle indices 1 and 2, respectively)
which is independent of the physical characteristics of the
situation. This
kind of permutation has no physical consequences and cannot be significant for
statistical mechanics.

At this point we conclude that if permutation is
understood as a physical interchange via trajectories in
space-time, then
permutations represent physically different possibilities, in the
sense of different physical processes that may be realized.
However, if permutation is understood differently, then
the permutations may not be
associated with any physical differences.

\section{Permutations in statistical mechanics}

Let us now take a closer look at the question of what kind of
permutations are relevant to statistical mechanics -- physical
exchanges with connecting trajectories, or simply swapping indices? Which kind of permutations determine the number of
microstates $W$?

Consider again two gas-filled chambers, each containing $N$
identical particles. Before the partition is removed the number of
available states per particle is $X$. After the partition has been
removed, the number of available states is $2X$. The reason is
that after the partition's removal it has become possible for the
particles to move to the other chamber. The doubling of the
number of available microstates thus expresses a physical freedom
that was not present before the partition was taken away.
Trajectories in space-time have become possible from the
particles' initial states to states in the other chamber. Particles from the left and right sides can physically exchange their states.

In contrast, even with the partition in place we could imagine the permutation of particles from the left and right
sides, but such permutations are not taken into account in the
calculation of the multiplicity. Nor do we consider permutations
with similar particles outside of the container. In
other words, the relevant kind of permutation is the physical
exchange, not the swapping of indices.

To completely justify the answer that accessibility via a real
physical process, associated with a connecting trajectory, is the determining
factor in the calculation of multiplicities, we would have to
consider the foundations of statistical mechanics.
An important approach in this area is
ergodic theory, in which the probability of a macrostate is
argued to be proportional to its multiplicity on the grounds that
the multiplicity is a measure of the time a system actually
spends in that part of phase space corresponding to the
macrostate in question. This argument makes sense only if the
microstates in this part of the phase space are actually
accessible by physical processes. Microstates that give rise to
the same macrostate but cannot be reached from the initial
microstate through the evolution of the system are irrelevant for
the macrostate's probability.

Although the original form of the ergodic hypothesis
(according to which all microstates are actually visited in a
relatively short time) has proven to be untenable, the basic idea that accessibility is the criterion for
the relevance of microstates is still essential. The multiplicities that occur in
more recent and sophisticated approaches to the foundations
of statistical mechanics are the same as those of the original
ergodic theory.

We therefore conclude that the multiplicity of a macrostate in classical statistical mechanics is given by the number of ways that
the macrostate can be reached by a physical process.
Permutations, corresponding to physical exchanges, represent
different and real physical possibilities. We must therefore not divide by
$N!$ when calculating multiplicities of macrostates of identical
classical particles.

\section{Empirical consequences of not dividing by $N!$}\label{empcons}

The foregoing argument for not dividing by $N!$, plus
the relation $S=k \ln W$, makes the statistical mechanical entropy
non-extensive. For example, after the removal of the partition the total
entropy is not twice the entropy each single chamber had before,
but is larger by the amount $2kN \ln2$.

There are three kinds of objections
against not dividing by $N!$ and the non-extensivity of the
statistical entropy that is its consequence: (1) it leads to incorrect
empirical predictions;\cite{schrodinger 1948, lin 1996, swendsen
2002} (2) it leads to a violation of the second law of
thermodynamics;\cite{wannier 1966, schroeder 2000, swendsen 2008}
and (3) it leads to a discrepancy with the thermodynamic entropy.

In this section we address the first of these objections. The
second and the third, the Gibbs paradox proper, are discussed
in Secs.~\ref{second law} and \ref{solution}.

In contrast to what is claimed by some authors,\cite{schrodinger
1948, lin 1996, swendsen 2002} dividing or not dividing the number
of microstates by $N!$ is irrelevant to the empirical predictions
of statistical mechanics for isolated systems. That is, although
there is no fundamental justification for dividing by $N!$ on the
basis of particle indistinguishability, a systematic division by
$N!$ for all particles of the same kind in mutually accessible
states has no empirical consequences. This conclusion follows because all
empirical predictions made by statistical mechanics rest
ultimately on the probabilities assigned to macrostates. These
probabilities are calculated with the help of the fundamental
assumption that in equilibrium the accessible microstates in an
isolated system are all equally probable, so that the probability
of a macrostate is its multiplicity divided by the total number of
microstates. In isolated systems no particles can move in or
out, so that the number $N$ of particles remains constant which
means that not only are all multiplicities decreased by the factor of
$1/N!$: the same happens to the total number of microstates if we
decide to systematically divide by $N!$. The probability of a
macrostate is therefore not affected, and whether
we divide the multiplicity by $N!$ or not makes no difference
for empirical predictions.

Statistical mechanics does not only describe systems
with a constant number of particles. Systems in which particles
can move in and out can also be treated, as is done with the
grand canonical ensemble, where the
probability distribution can be derived by considering a system
with varying particle numbers as part of a larger particle
reservoir. This reservoir is again isolated, and the fundamental
assumption is now applied to this larger system. Therefore
our conclusion that the empirical predictions remain the same,
whether we systematically divide by $N!$ or not, also applies to
systems in which particles are exchanged between subsystems (in
agreement with what was argued in Ref.~\onlinecite{van kampen 1984}).

Particle numbers can also change by chemical reactions, where
the number of molecules of a certain kind need not be
constant, even if the system is isolated. For such a
system division by the number of permutations of the elementary
constituents (in this case the atoms) does not matter for the
empirical results because the numbers
of atoms from which the molecules are composed remain
constant, as discussed in Ref.~\onlinecite{ehrenfest 1920}.

To illustrate that the factor $1/N!$ does not matter for empirical
predictions, we consider the example given in Ref.~\onlinecite{swendsen 2002}.\cite{discussion article swendsen} Two containers, one of volume $V_1$ and the other of volume
$V_2$, are separated by a wall. Both contain the same kind of
ideal gas, consisting of identical classical particles. Now a hole
is poked in the wall and the particles can move to the other
container. After some time equilibrium is reached. What is the number of particles in the volume $V_1$ in
equilibrium? From standard classical statistical mechanics, with
the factor $1/N!$ included in the multiplicity, the empirical
prediction follows that $N_1=N V_1/(V_1+V_2)$, where $N_1$ is the
number of particles in $V_1$ and $N$ is the total number of
particles. According to Ref.~\onlinecite{swendsen 2002}, if the
factor $1/N!$ were not included, an incorrect empirical prediction would
be obtained if $V_1$ is not equal to $V_2$: all particles would go to the larger of
the two volumes.

However, as we shall now demonstrate, consistently taking into
account the particles' distinguishability does lead to
the correct predictions using $S=k\ln W$. In Ref.~\onlinecite{swendsen 2002} empirical predictions were obtained by
calculating in what macrostate the entropy is maximal, and we take
the same approach here.

Consider a macrostate of the ideal gas, defined by the observable
$N_1$. As a constraint
we take that the total number of particles is $N$, so
that $N-N_1$ particles are in volume $V_2$. The multiplicity is
given by
\begin{equation}\label{mult}
W=c\,V_1^{N_1}\,V_2^{N-N_1}\frac{N!}{N_1!(N-N_1)!},
\end{equation}
where $c$ is a factor independent of the volume and the number of
particles. The factor $N!/(N_1!(N-N_1)!)$ is included because it
is the number of ways $N_1$ distinguishable particles can be
chosen out of $N$ particles. With $S=k\ln W$ and Stirling's
approximation $\ln N!\approx N \ln N-N$, we find for the entropy
\begin{eqnarray}
S&=&kN_1\ln V_1 +k(N-N_1)\ln V_2 \nonumber\\
&&{}+kN\ln N-kN_1\ln N_1 -k(N-N_1)\ln (N-N_1)+k\ln c. \label{this}
\end{eqnarray}
In equilibrium $S$ is a maximum, so $N_1$ in equilibrium
can be found by solving $\partial S/\partial N_1 = 0$.
The solution is $N_1=N V_1/(V_1+V_2)$, which is the same empirical
prediction as obtained with the factor $1/N!$ included.

Note how the non-extensivity of the entropy $S=k\ln W$ proves
essential for a consistent use of this expression for the entropy.
When a hole is made in the wall, particles in volume $V_1$ may exchange their positions with particles in the other
volume, so the multiplicity increases as indicated in Eq.~(\ref{mult}). That means that the entropy becomes larger than the
sum of the two original entropies of $V_1$ and $V_2$, just as in
the case of the Gibbs paradox.\cite{non-extensivity}

\section{Second law of thermodynamics}\label{second law}

The second alleged problem is that the non-extensivity of the
entropy leads to a violation of the second law of thermodynamics.
\cite{wannier 1966, schroeder 2000, swendsen 2008} Consider again the
two chambers after the partition has been
removed. When the partition is returned, the entropy decreases
by $2kN \ln2$. But the second law insists that entropy cannot
decrease!

An important issue here is the nature of the second law in its statistical version: it will turn out that to solve the just-mentioned problem we need to recognize
that the second law in thermodynamics is not exactly the same as
the second law in statistical mechanics.

In thermodynamics the second law says that the entropy does not
decrease if there is no heat transfer to or from the environment. For the case of the replacement of the partition there is no such
heat transfer, so the thermodynamic entropy should not decrease.

In contrast, in statistical mechanics the second law expresses a probability consideration: equilibrium macrostates possess a probability vastly greater than that of non-equilibrium
states, and therefore it is enormously probable that the system
will reach equilibrium and subsequently stay in equilibrium. The
number of microstates corresponding to the equilibrium macrostate
is virtually equal to the total number of accessible microstates
$W$, which is the justification for the use of
$S=k\ln W$ as the entropy belonging to the equilibrium state. That
the statistical mechanical entropy $S$, with $S=k \ln W$, does not
decrease is a direct expression of the dominance of the most
probable macrostate. Accordingly, the second law of statistical
mechanics states that the entropy of a system goes, with
overwhelming probability, to the maximum value compatible with the
total number of accessible microstates, and will
not decrease as long as this number of accessible states does not decrease. The statistical second law tells us, for instance, that
without intervention from outside not all particles will move to
the same corner of the container. But it does not tell us how $W$
may change when the system is manipulated from outside, for
example by placing a partition in it.

\section{Solution of the Gibbs paradox}\label{solution}

The paradox is that upon removal of the partition between the two
containers the entropy increases according to classical statistical
mechanics, whereas it remains the same in thermodynamics. What we
have seen is that for classical particles
the entropy should really increase in statistical mechanics, in
the sense that the number of accessible microstates $W$ does
increase. This increase reflects the redefinition of the system,
and of $W$ that results from manipulation from outside (the
removal of the partition). The statistical second law only plays a
role here to the extent that it says that the system will assume
an equilibrium state which is compatible with this new and
increased $W$. In principle, we could empirically verify
that the number of microstates has
actually increased by following the paths of individual particles
(we could in this way give empirical content to the existence
of the entropy of mixing.\cite{dieks 2010}) But this verification would require
measurements on the microscopic level that would lead us outside
the domains of thermodynamics and statistical
mechanical predictions (in which the precise microstate is assumed
to be immaterial). As we have seen in Sec.~\ref{empcons}, as long as we remain within the realm of the
usual macroscopic measurements, the increase of the entropy,
for gases of the same kind, will not lead to any new empirical consequences.

Is there a problem here? Can we not just accept that the entropy
changes in statistical mechanics and remains the same in
thermodynamics? Yes, we can. We only have to take into account that the entropy is defined
differently in statistical mechanics than in thermodynamics. In
statistical mechanics the existence of a micro-description is
taken into account as a matter of principle, whereas in
thermodynamics this same micro-description is excluded from the
start. As we have seen, the consequence is an increase of entropy according
to statistical mechanics, but its direct empirical significance is only on the micro-level and does not lead to any new
predictions for macroscopic measurements. The difference between
the statistical mechanical and the thermodynamic definition of
entropy makes it understandable that the values of
entropy changes in statistical mechanics are sometimes different
from those in thermodynamics.

There are no conflicts with the second law here. According to
both statistical mechanics and thermodynamics, the second
law is obeyed perfectly. The statistical mechanical account is
that the number of microstates increases through intervention from
outside (the removal of the partition), and the statistical second law
tells us that the system will subsequently make maximum use of
the increased area in phase space. Therefore, the prediction is
that the system will exhibit equilibrium values of its macroscopic
quantities, so that on the macro-level nothing changes. In
thermodynamics this ``fine-grained'' consideration plays no role:
nothing changes and according to the thermodynamic
second law the entropy remains constant. Both descriptions are
valid within their own context and lead to the same predictions.

The same type of argument can be used if we consider what happens
when the partition is replaced. According to thermodynamics
nothing happens to the entropy, because there is no exchange of heat.
According to statistical physics, there is a decrease of
$W$ and thus of the entropy. But this decrease does not mean a violation
of the second law! The decrease in $W$ is a consequence of the
redefinition of the accessible phase space caused by
intervention in the system. The statistical second law remains
valid, and predicts that the two subsystems will fully
make use of their phase space possibilities (which in this case
means that equilibrium will be maintained).

From a pragmatic point of view it is useful if
thermodynamics and statistical mechanics give exactly the same
entropy values. We can achieve this equivalence by introducing a new entropy
definition in statistical mechanics: Replace $S = k \ln W$ by $S =
k \ln (W/N!)$. For systems in which $N$ is constant this new definition makes no
difference for any empirical predictions, because it only adds a
constant number to the entropy value. For the Gibbs case, in
which $N$ does change, this adaptation of the definition
leads to the disappearance of the mixing term and
therefore to extensivity of the entropy in the limit $N \to \infty$. In this way agreement with
thermodynamics is obtained.

It is important to realize that the introduction of this ``reduced
entropy'' (as it is called by Cheng\cite{cheng 2009}), is motivated by the wish to preserve this agreement: its justification is in the reproduction of thermodynamic entropy values under all conditions, even though the basic assumptions of statistical physics are different from those of thermodynamics. It is certainly reasonable to
change the definition of the statistical mechanical entropy to achieve
agreement with thermodynamics. There is nothing
wrong with the reduced entropy if it is interpreted from this pragmatic point of view. But it would be a mistake to interpret
the division by particle number factorials as a necessary consequence of a fundamental
feature of microscopic physical reality, namely the indistinguishability of particles of the same kind.

\section{Distinguishable identical quantum particles}

In contrast to classical particles, identical quantum particles are usually assumed to be indistinguishable as a matter of principle. This fundamental
indistinguishability is supposed to be the result of the quantum mechanical
(anti-)symmetrization postulates, where permutations of the indices in the quantum mechanical many-particle formalism leave a state either invariant (for bosons)
or change its sign (for fermions).

As we have argued, the permutations that are relevant
to statistical mechanics are not simply permutations of indices, but rather processes by which the permuted state is
connected to the initial state by trajectories in space-time. Quantum mechanics allows situations in which definite
particle trajectories through space-time can be defined. When the
individual one-particle wave packets in the many-particle state
have small spatial extensions and do not overlap, we can identify
them over time (and by Ehrenfest's theorem they approximately
follow the paths of classical particles). In the classical limit
of quantum mechanics such narrow wave packets come to represent
the particles we know from classical physics. Permutation of
trajectories of such wave packets in the physical sense discussed
before produces a different state, just as in classical physics.
However, we have all been taught that quantum particles of the
same kind are fundamentally indistinguishable -- so how can it be
true that it may make sense to speak about a physical interchange
of them?

The confusion results from an ambiguity in how the particle
concept is used in quantum mechanics. Sometimes the indices in the
many-particle formalism are taken to refer to particles so that
everything is symmetrical and all particles are in exactly the same state (as follows from taking partial traces to define the one-particle states). But there is also another use of
the particle concept in which individual wave packets are taken to
represent single particles. It is this second way of thinking
about particles, and not the first one, that in the classical
limit gives the particles we know from classical physics.
\cite{dieks 2010, dieks 2008, dieks to appear} Classical particles
emerge from quantum mechanics when wave
packets become localized in the classical limit. It is only then that the classical particle
concept becomes applicable (see Ref.~\onlinecite{dieks to appear}
for an extensive discussion).

Narrow wave packets can follow trajectories through space-time, and as
long as they remain separated from each other, these wave packets remain as
distinguishable as classical particles. When a quantum particle
represented by such a wave packet moves from $x_1$ to $x_2$ and
another particle of the same kind moves from $x_2$ to $x_1$, then the
final state is different from the initial state just as in our
earlier discussion on classical particles.

As we have argued, the Gibbs paradox can be resolved without
invoking the indistinguishability of particles. We see that
quantum theory does not change the relevant characteristics of the
situation, so that quantum theory is irrelevant to
the resolution of the paradox. It does not matter if we
switch to the quantum description of the gases we have considered
in the context of the paradox. If the particles are
(approximately) localized, permutation produces a different microstate just as in classical
mechanics.

It is interesting to compare our arguments with those of Saunders,
\cite{saunders 2006} who regards even classical particles as
indistinguishable. According to Saunders there is one reason why
classical particles are generally assumed to be distinguishable:
``$\ldots$ we surely \emph{can} single out classical particles
uniquely, by reference to their trajectories. But there is a key
objection to this line of thinking: \emph{so can quantum
particles, at least in certain circumstances, be distinguished by
their states}.'' We agree with this conclusion, but we do not see it
as an objection. Saunders' conclusion is that
``indistinguishability (permutability, invariance under
permutations) makes just as much sense classically as it does in
quantum mechanics.'' We say the opposite:
distinguishability often makes just as much sense for quantum
particles as for classical particles.

\section{Conclusion}

The Gibbs paradox should not be interpreted as providing an
argument for the indistinguishability of identical classical
particles. Classical particles
can be distinguished by their different positions and their
different trajectories through space-time. Nor does the Gibbs paradox needs quantum mechanics for its solution:
identical quantum particles can be as
distinguishable, in the relevant sense, as classical particles.

One way of resolving the Gibbs paradox is by recognizing that
entropy and the second law have slightly different meanings in
statistical mechanics and in thermodynamics. In thermodynamics the
entropy has to be extensive; but this is not necessarily so in
statistical mechanics. It is possible, and even
advisable, to make the statistical mechanical entropy extensive by
introducing a new entropy definition (the reduced entropy) that does not change
predictions as long as these predictions remain within the realm
of thermodynamics. But this convention has nothing to do with any
fundamental indistinguishability of
identical particles.


\begin{thebibliography}{99}
\bibitem{number of states}
For simplicity, we assume here that the number of
available states is finite, though very large compared to
the number of particles.
\end{thebibliography}
\end{document}